\documentclass{raa}            
\usepackage{graphicx}
\usepackage{natbib}

\begin{document}

   \title{Winding sense of galaxies around the Local Supercluster
}

 \volnopage{{\bf 2010} Vol.{\bf XX} No.{\bf X}, 000--000}
   \setcounter{page}{1}

   \author{B. Aryal
      \inst{1,2}
   }

   \institute{Central Department of Physics, Tribhuvan University, Kirtipur, Nepal {\it binil.aryal@uibk.ac.at}\\
        \and Institut of Astro- and Particle Physics, Innsbruck University, Technikerstrasse 25, A-6020 Innsbruck, Austria\\
\vs \no
   {\small Received [2010] [January] [14]; accepted [2010] [October] [23] }
}

\abstract{ We present an analysis of the winding sense (S and
Z-shaped) of 1\,621 field galaxies that have radial velocity
between 3\,000 km s$^{-1}$ to 5\,000 km s$^{-1}$. The preferred
alignments of S- and Z-shaped galaxies are studied using
chi-square, auto-correlation and the Fourier tests. We classified
total galaxies into 32 subsamples and noticed a good agreement
between the position angle (PA) distribution of S- and Z-shaped
galaxies. The homogeneous distribution of the S- and Z-shaped
galaxies is noticed for the late-type spirals (Sc, Scd, Sd and Sm)
than that of the early-types (Sa, Sab, Sb and Sbc). A significant
dominance of S-mode galaxies is noticed in the barred spirals. A
random alignment is noticed in the PA-distribution of Z- and
S-mode spirals. In addition, homogeneous distribution of the S-
and Z-shaped galaxies is found to be invariant under the global
expansion. The PA-distribution of the total S-mode galaxies is
found to be random, whereas preferred alignment is noticed for the
total Z-mode galaxies. It is found that the galactic planes of
Z-mode galaxies tend to lie in the equatorial plane.
 }

   \authorrunning{B. Aryal}            
   \titlerunning{Winding Sense of Galaxies }  
   \maketitle


%
%
\section{Introduction}           
\label{sect:intro}

Differential rotation in a galaxy's disc generate density waves in
the disc, leading to spiral arms. According to gravitational
theory, the spiral arms born as leading and subsequently transform
to trailing modes. With the passage of time, the spiral pattern
deteriorates gradually by the differential rotation of the
equatorial plane of the galaxy, but the bar structure persists for
a long time (Oort 1970a). This structure can again regenerate
spiral pattern in the outer region. Thus, a close relation between
the origin of the arms in the spirals and barred spirals can not
be denied (Oort 1970b).

Land et al. (2008) studied the distribution of projected spin
vectors of $\sim$ 37\,000 spiral galaxies taken from Solon Dizital
Sky Survey. They did not notice any evidence for overall preferred
handedness of Universe. In a similar study, Longo (2007) found
evidence for a preferred axis. Sugai \& Iye (1995) used statistics
and studied the winding sense of galaxies (S- and Z-patterns) in
9\,825 spirals. No significant dominance from a random
distribution is noticed. Aryal \& Saurer (2005) studied the
spatial orientations of spin vectors of 4\,073 galaxies in the
Local Supercluster. No preferred alignment is noticed for the
total sample. These results hint that the distribution of angular
momentum of galaxies is entirely random in two- (S- and Z-shaped)
and three-dimensional (spin vector) analysis provided the database
is rich.

In order to understand true structural modes (leading or trailing)
of spiral galaxies, we need to know the direction of the spiral
pattern (S- or Z-patterns), the approaching and receding sides and
the near and far parts, since galaxies are commonly inclined in
space to the line of the sight. The S and Z-patterns can be
determined from the image of the galaxy. Similarly, the
approaching and receding sides can be defined if spectroscopy data
on rotation is available. The third one is fairly hard to
established. For this, Pasha (1985) used `tilt' criteria and
studied the sense of winding of the arms in 132 spirals. He found
107 spirals to have trailing arm. Thomasson et al. (1989) studied
theoretically and performed $N$-body simulations in order to
understand the formation of spiral structures in retrograde galaxy
encounters. Interestingly, they noticed the importance of halo
mass. They concluded that the spirals having halos with masses
larger than the disk mass exhibit leading pattern. Thus, the
makeup of galactic haloes is important to cosmology in order to
understand the evolution of galaxies.

By considering the group of transformations acting on the
configuration space, Capozziello \& Lattanzi (2006) predicted that
the progressive loss of inhomogeneity in the S- and Z-shaped
galaxies might have some connection with the
rotationally-supported (spirals, barred spirals) and randomized
stellar systems (lenticulars, ellipticals).

The  preferred alignments of galaxies can be an indicator of
initial conditions when galaxies and clusters formed provided the
angular momenta of galaxies have not been altered too much since
their formation. A useful property of galaxies in clusters for
which theories make different predictions is the angular momentum
distribution. The `Pancake model' by Doroshkevich
\cite{Doroshkevich}, the `Hierarchy model' by Peebles
\cite{Peebles} and, the `Primordial vorticity model' by Ozernoy
\cite{Ozernoy} predict different scenarios concerning the
formation of large-scale structure. Thus, the study of galaxy
orientation has the potential to yield important information
regarding the formation and evolution of cosmic structures. In
this work, we present an analysis of winding sense and preferred
alignments of galaxies that have radial velocity (RV) 3\,000 km
s$^{-1}$ to 5\,000 km s$^{-1}$. These are field galaxies. We
intend to study the importance of winding sense in order to
understand the true structural modes (i.e., leading and trailing
arm) of the galaxy. We expect to study the following: (1) Are the
distribution of S- and Z-shaped galaxies homogeneous in the field?
(2) Is there any correlation between the preferred alignment and
the winding sense of galaxies? (3) Does radial velocity dependence
exist concerning winding sense of galaxies? and finally, (4) What
can we say about the distribution of true structural modes (i.e.,
leading or trailing arm) of galaxies in the large scale structure?

This paper is organized as follows: in Sect. 2 we describe the
method of data reduction. In Sect. 3 we give a brief account of
the methods and the statistics used. Finally, a discussion of the
statistical results and the conclusions are presented in Sects. 4
and 5.


\section{The sample: data reduction}
\label{sect:Obs}

\begin{figure}
\vspace{0.0cm}
      \centering
       \includegraphics[height=3.8cm]{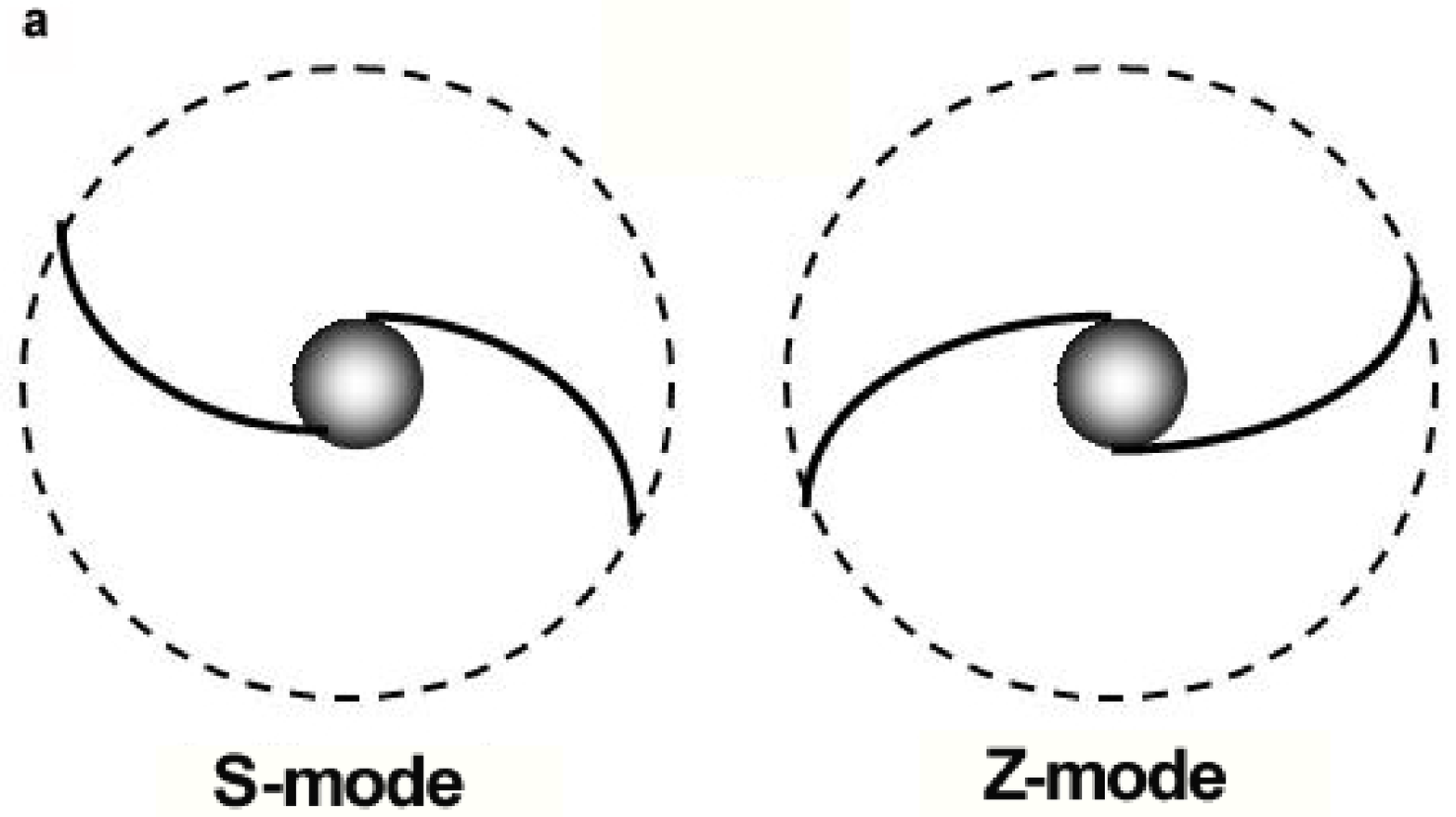}
       \includegraphics[height=3.8cm]{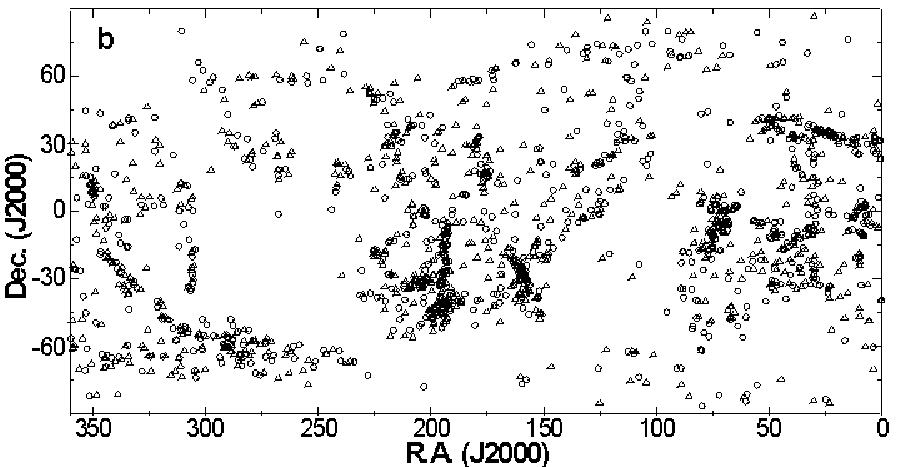}
       \includegraphics[height=3.4cm]{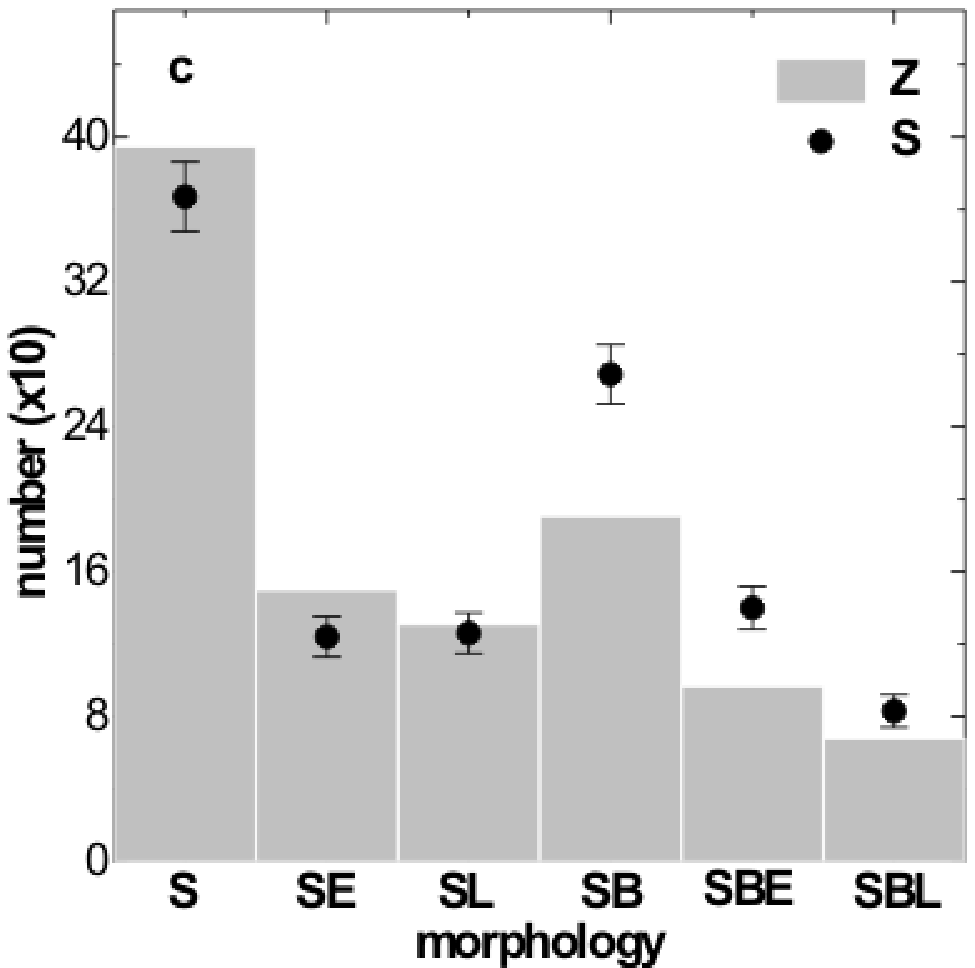}
       \includegraphics[height=3.4cm]{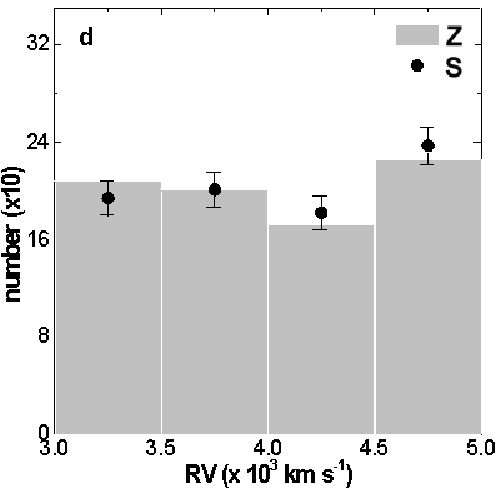}
       \includegraphics[height=3.4cm]{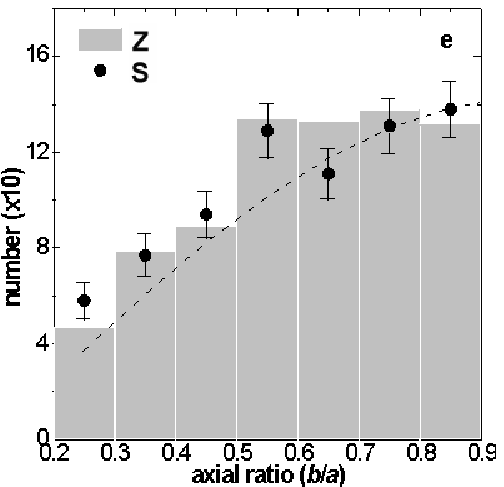}
       \includegraphics[height=3.4cm]{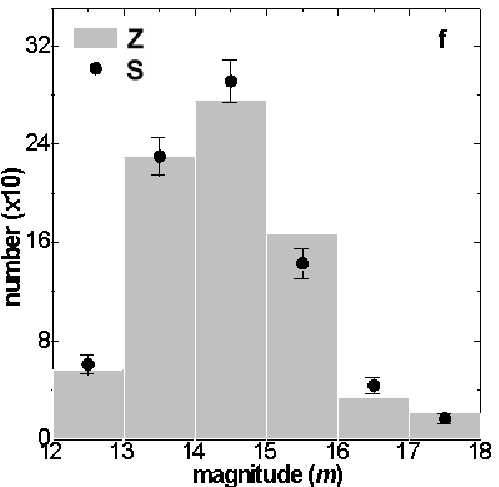}
      \caption[]{(a) A sketch representing the winding sense (S or Z) of the galaxy. (b) All-sky distribution
      of Z-mode ($\triangle$) and S-mode ($\circ$) galaxies
      that have RVs in the range 3\,000 km s$^{-1}$ to 5\,000 km s$^{-1}$.
      The morphology (c), radial velocity (d), axial ratio (e) and the
      magnitude (f) distribution of Z and S-mode galaxies in our database.
      The statistical $\pm$1$\sigma$ error bars are shown for the S-mode ($\bullet$) subsample.
      The dashed line (e) represents the expected distribution. }
\end{figure}

Eighteen catalogues were used for the data compilation. A list of
the catalogues and their references are given in Table 1. The
abbreviations given in the first column of Table 1 are as follows:
NGC - New General Catalogue, UGC - Uppsala General Catalogue of
Galaxies, ESO - ESO/Uppsala Survey of the ESO (B) Atlas, IC -
Index Catalogue, MCG - Morphological Galaxy Catalogue, UGCA -
Uppsala obs. General Catalogue, Addendum, CGCG - Catalogue of
Galaxies and Clusters of Galaxies, KUG - Kiso Ultraviolet Galaxy
Catalogue, MRK - Markarian Galaxy Catalogue, MESSIER - Catalogue
des nebuleuses et des amas d'etoiles, BCG - Brandner+Grebel+Chu
Catalogue, LSBG - Low Surface Brightness Galaxies, SBS - Second
Byurakan Survey, LCRS - Las Companas Red Shift Survey, DDO - David
Dunlap Observatory Publications, IRAS - Infrared Astronomical
Satellite, SGC - Southern Galaxy Catalogue and UM - University of
Michigan: Curtis Schmidt-thin prism survey for extragalactic
emission-line objects: List I-V.

The NASA/IPAC extragalactic database (NED,
http://nedwww.ipac.caltech.\\edu/) was used to compile these
catalogues. The main editing process was as follows: first,
galaxies having RVs in the range 3\,000 km s$^{-1}$ to 5\,000 km
s$^{-1}$ were collected. We downloaded the image of all these
galaxies from NED in FITS format. The second step was to compile
the morphology of these galaxies from the catalog. A galaxy with
doubtful morphology (eg., `S?', `S0' or `Sa') is omitted. Finally,
the position angles (PAs) of galaxies were added from the UGC,
ESO, and Third Reference Catalogue of Bright Galaxies (de
Vaucouleurs et al. 1991).

\begin{table}
\vspace{0.0cm} \caption{The list of catalogues used for the data
compilation. The first column lists the abbreviation of the
catalogue. The second column gives the total number of galaxies.
The references are listed in the last column.}
$$
\begin{array}{p{0.12\linewidth}rll}
\hline \noalign{\smallskip}
Catalogue  & $N$ &   $References$          \\
\noalign{\smallskip} \hline \noalign{\smallskip}
NGC       & 623  &    $Dreyer (1895, 1908) $    \\
UGC       & 276  &    $Nilson (1973)$           \\
ESO       & 123  &    $Lauberts (1982)$         \\
IC        & 93   &    $Dreyer (1895, 1908) $    \\
MCG       & 88   &    $Vorontsov-Vel'Yaminov et al.$ \\
          &      &    $(1962-74)$\\
UGCA      & 84   &    $Nilson (1974)$           \\
CGCG      & 75   &    $Zwicky et al. (1961-68)$ \\
KUG       & 44   &    $Takase (1980-2000)$      \\
MRK       & 40   &    $Markarian (1967)$        \\
MESSIER   & 41   &    $Messier (1784)$          \\
BCG       & 28   &    $Brandner et al. (2000)$  \\
LSBG      & 23   &    $Impey et al. (1996)$     \\
SBS       & 17   &    $Markarian et al. (1983)$ \\
LCRS      & 16   &    $Shectman et al. (1996)$  \\
DDO       & 15   &    $Bergh (1959, 1966)$      \\
IRAS      & 15   &    $Infrared Astronomical Satellite$  \\
          &      &    $(1983)$\\
SGC       & 10   &    $Corwin et al. (1985)$    \\
UM        & 10   &    $MacAlpine et al.$  \\
          &      &    $(1977a,b,c; 1978; 1981)$ \\
\noalign{\smallskip} \hline
\end{array}
$$
\end{table}

There were two clusters Abell 0426 ($\alpha$(J2000) = 03$^{\rm
h}$18$^{\rm m}$ 36.4$^{\rm s}$, $\delta$(J2000) =
+41$^{\circ}$30'54'') and Abell 3627 ($\alpha$(J2000) = 16$^{\rm
h}$15$^{\rm m}$32.8$^{\rm s}$, $\delta$(J2000) =
--60$^{\circ}$54'30'') in our region. These clusters have mean RVs
5\,366 km s$^{-1}$ (75 $\pm$ 5 Mpc) and 4\,881 km s$^{-1}$ (63
$\pm$ 4 Mpc), respectively (Abell, Corwin \& Olowin 1989, Struble
\& Rodd 1999). We removed the galaxies belong to the cluster Abell
0426 using the catalog established by Brunzendorf \& Meusinger
(1999). For the cluster Abell 3627 galaxies, we used Photometric
Atlas of Northern Bright Galaxies (Kodaira, Okamura \& Ichikawa
1990) and Uppsala Galaxy Catalogue (Nilson 1973). There were 174
galaxies belongs to these clusters in our database. We remove
these galaxies.

The RVs were compiled from Las Campanas Redshift Survey (Shectman
1996). The PAs and the diameters of galaxies were added from the
Uppsala Galaxy Catalogue (Nilson 1973), Uppsala obs. General
Catalogue, Addendum (Nilson 1974), Photometric Atlas of Northern
Bright Galaxies (Kodaira, Okamura, \& Ichikawa 1990), ESO/Uppsala
Survey of the European Southern Observatory (Lauberts 1982),
Southern Galaxy Catalogue (Corwin et al. 1985) and Third Reference
Catalogue of Bright Galaxies (de Vaucouleurs et al. 1991).

In the NED, 6\,493 galaxies having RVs 3\,000 km s$^{-1}$ to
5\,000 km s$^{-1}$ were listed until the cutoff date.
Morphological information was given in the catalogues for 3\,276
(50\%) galaxies. We visually inspected all these galaxies using
ALADIN2.5 software.

The arm patterns (S- or Z-type) of the galaxies were studied
visually by the author in order to maintain homogeneity. The
contour maps of the galaxies were studied in order to identify
their structural modes. For this, we used ALADIN2.5 software. The
Z-mode is one whose outer tip points towards the anti-clockwise
direction (see Fig. 1a). Similarly, the outer tip of S-mode
directs in the clockwise direction. These two patterns are
obviously the two dimensional projections of three dimensional
galaxy. The re-examination of the S- and Z-modes using MIDAS
software resulted the rejection of more than 17\% of the objects.
These rejected galaxies were nearly edge-on spiral and barred
spiral galaxies. As expected, it was relatively easier to identify
the structural modes of nearly face-on than that of nearly edge-on
galaxies.

In this way, we compiled a database of 1\,621 galaxies showing
either S- or Z-structural mode. There were 807 Z-mode and 821
S-mode galaxies in our database. All sky distribution of Z- and
S-mode galaxies is shown in Fig. 1b. The symbols ``$\circ$" and
``$\triangle$" represent the positions of the S-mode and Z-mode
galaxies, respectively. Several groups and aggregations of the
galaxies can be seen in the figure.

The morphology, radial velocity, axial ratio and the magnitude
distributions of S- and Z-mode galaxies are shown in Figs.
1c,d,e,f. The spirals (47\%) dominate our database (Fig. 1c).
However, a significant dominance of S-modes are noticed in the
barred spirals whereas a weak dominance of Z-modes are found in
the spirals. The population of galaxies in the RV distribution
($\Delta$RV = 5\,00 km s$^{-1}$) were nearly equal (Fig. 1d). The
axial ratio distribution shows a good agreement with the expected
{\it cosine} curve in the limit 0.2 $<$ $b$/$a$ $\leq$ 0.9 (Fig.
1e). The values of absolute magnitude lie between 13.0 and 16.0
for 82\% galaxies in our database (Fig. 1f).

We classified the database into 32 subsamples for both the S- and
Z-modes on the basis of the morphology, radial velocity, area and
the group of the galaxies. The galaxies with doubtful morphology
are omitted in the spiral and barred spiral subsamples. The total
number of early- and late-type spirals or barred spirals is much
less than that of the total spiral or total barred spirals. It is
because of the fact that the galaxy with incomplete morphology,
say, simply `S' or `SB' can not be included in the subsamples. In
other words, the galaxy with morphology Sa, Sab, Sb, Sbc are
included in the early spirals whereas the galaxies with morphology
Sc, Scd, Sd and Sdm are classified as late type spirals. The
galaxies having morphology other than Sa, Sab, Sb, Sbc, Sc, Scd,
Sd and Sm can not be included in the early and late subsamples. A
statistical study of these subsamples are given in Table 1 and
discussed in Sect. 4.1.

\section{Method of analysis}
\label{sect:data}

Basic statistics is used to study the dominance of Z- or S-mode
galaxies. At first, morphology and RV dependence are studied.
Secondly, sky is divided into 16 equal parts in order to observe
deviation from the homogeneity. Several galaxy groups are
identified in the all-sky map where the structural dominance are
noticed. Finally, we study the dominance of Z- or S-mode galaxies
in these groups.

We assume isotropic distribution as a theoretical reference and
studied the equatorial PA-distribution in the total sample and
subsamples. In order to measure the deviation from isotropic
distribution we have carried out three statistical tests:
chi-square, auto-correlation and the Fourier.

We set the chi-square probability P($>\chi^2$) = 0.050 as the
critical value to discriminate isotropy from anisotropy, this
corresponds to a deviation from isotropy at the 2$\sigma$ level
(Godlowski 1993). Auto correlation test takes account the
correlation between the number of galaxies in adjoining angular
bins. We expect, auto correlation coefficient C$\rightarrow$0 for
an isotropic distribution. The critical limit is the standard
deviation of the correlation coefficient C.

If the deviation from isotropy is only slowly varying with angles
(in our case: PA) the Fourier test can be applied (Godlowski
1993). A method of expanding a function by expressing it as an
infinite series of periodic functions ({\it sine} and {\it
cosine}) is called Fourier series. Let $N$ denote the total number
of solutions for galaxies in the sample, $N$$_{k}$ the number of
solutions in the k$^{th}$ bin, $N$$_{0}$ the mean number of
solutions per bin, and $N$$_{0k}$ the expected number of solutions
in the k$^{th}$ bin. Then the Fourier series is given by (taking
first order Fourier mode),

\begin{equation}
\begin{array}{l}
N_{k} = N_{k}(1+ \Delta_{11} \cos 2\beta_{k}+ \Delta_{21} \sin 2\beta_{k}+ ......)\\
\end{array}
\end{equation}
Here the angle {\bf $\beta$$_{k}$} represents the polar angle in
the k$^{th}$ bin. The Fourier coefficients  $\Delta_{11}$ and
$\Delta_{21}$ are the parameters of the distributions. We obtain
the following expressions for the Fourier coefficients
$\Delta_{11}$ and $\Delta_{21}$,
\begin{equation}
\begin{array}{l}
\Delta_{11} = \sum (N_{k}-N_{0k}) \cos 2\beta_{k} / \sum N_{0k} \cos^2 2\beta_{k} \\
\end{array}
\end{equation}
\begin{equation}
\begin{array}{l}
\Delta_{21} = \sum (N_{k}-N_{0k}) \sin 2\beta_{k} / \sum N_{0k} \sin^2 2\beta_{k} \\
\end{array}
\end{equation}
The standard deviations  ($\sigma$($\Delta_{11}$)) and
($\sigma$($\Delta_{21}$)) can be estimated using the expressions,
\begin{equation}
\begin{array}{l}
\sigma (\Delta_{11}) = (\sum N_{0k} \cos^2 2\beta_{k})^{-1/2} \\
\end{array}
\end{equation}
\begin{equation}
\begin{array}{l}
\sigma (\Delta_{21}) = (\sum N_{0k} \sin^2 2\beta_{k})^{-1/2} \\
\end{array}
\end{equation}
The probability that the amplitude
\begin{equation}
\begin{array}{l}
\Delta_{1} = (\Delta_{11}^2 + \Delta_{21}^2)^{1/2} \\
\end{array}
\end{equation}
greater than a certain chosen value is given by the formula
\begin{equation}
\begin{array}{l}
P(>\Delta_{1}) = \exp(-nN_{0}\Delta_{1}^2/4) \\
\end{array}
\end{equation}
with standard deviation
\begin{equation}
\begin{array}{l}
\sigma (\Delta_{1}) = (2/nN_{0})^{1/2} \\
\end{array}
\end{equation}

The Fourier coefficient $\Delta_{11}$ gives the direction of
departure from isotropy. The first order Fourier probability
function $P$($>$$\Delta_{1}$) estimates whether (smaller value of
$P$($>$$\Delta_{1}$) or not (higher value of $P$($>$$\Delta_{1}$)
a pronounced preferred orientation occurs in the sample.

\section{Results}

First we present the statistical result concerning the
distribution of Z- and S-mode galaxies in the total sample and
subsamples. Second, we study the distribution of Z- and S-mode
galaxies in the unit area of the sky and in the groups. Then, the
equatorial PA-distribution of galaxies in the total sample and
subsamples are discussed. At the end, a general discussion and a
comparison with the previous results will be presented.

\subsection{Distribution of Z and S mode galaxies}

A statistical comparison between the total sample and subsamples
of the Z- and S-modes of galaxies is given in Table 2. Fig. 2
shows this comparison in the histogram. The $\Delta$(\%) in Table
1 and Fig. 2 represent the percentage difference between the
number Z- and S-mode galaxies. We studied the standard deviation
of the major diameters ($a$) of galaxies in the total sample and
subsamples for both the Z- and S-modes. In Table 2,
$\Delta(a\,$sde$)$ represents the difference between the standard
deviation of the major diameters of Z- and S-mode galaxies.

An insignificant difference (0.4\% $\pm$ 0.2\%) between the total
number of Z- and S-mode galaxies are found (Table 2). The
difference between the standard deviation of the major diameters
($\Delta(a\,$sde$)$) of the Z- and S-mode galaxies is found less
than 0.019 (eighth column, Table 2). Interestingly, the sum of the
major diameters of total Z- and S-mode galaxies coincide. This
result suggests the homogeneous distribution of Z- and S-mode
field galaxies that have RV in the range 3\,000 km s$^{-1}$ to
4\,000 km s$^{-1}$.

In Fig. 2, the slanting-line (grey-shaded) region corresponds to
the region showing $\leq$ 10\% (5\%) $\Delta$ value. Almost all
subsamples lie in this region, suggesting the homogeneous
distribution of Z- and S-mode galaxies within 10\% error limit.
Now, we present the distribution Z- and S-mode galaxies in the
subsamples classified according as their morphology, RVs, area and
the groups below.

\subsubsection{Morphology}

In the spirals, Z-mode galaxies are found 3.7\% ($\pm$1.8\%) more
than that of S-mode. The homogeneous distribution of Z- and
S-modes is found for the late-type spirals (Sc, Scd, Sd and Sm)
than that of early-type (Sa, Sab, Sb and Sbc): $\Delta$ value
turned out to be 9.5\% ($\pm$4.8\%) and 1.8\% ($\pm$1.0\%) for
early- and late-types (Table 2). Thus, no preferred winding
pattern is noticed in the late-type spirals than that of
early-types.

\begin{figure}
\vspace{0.0cm}
      \centering
      \includegraphics[height=3.55cm]{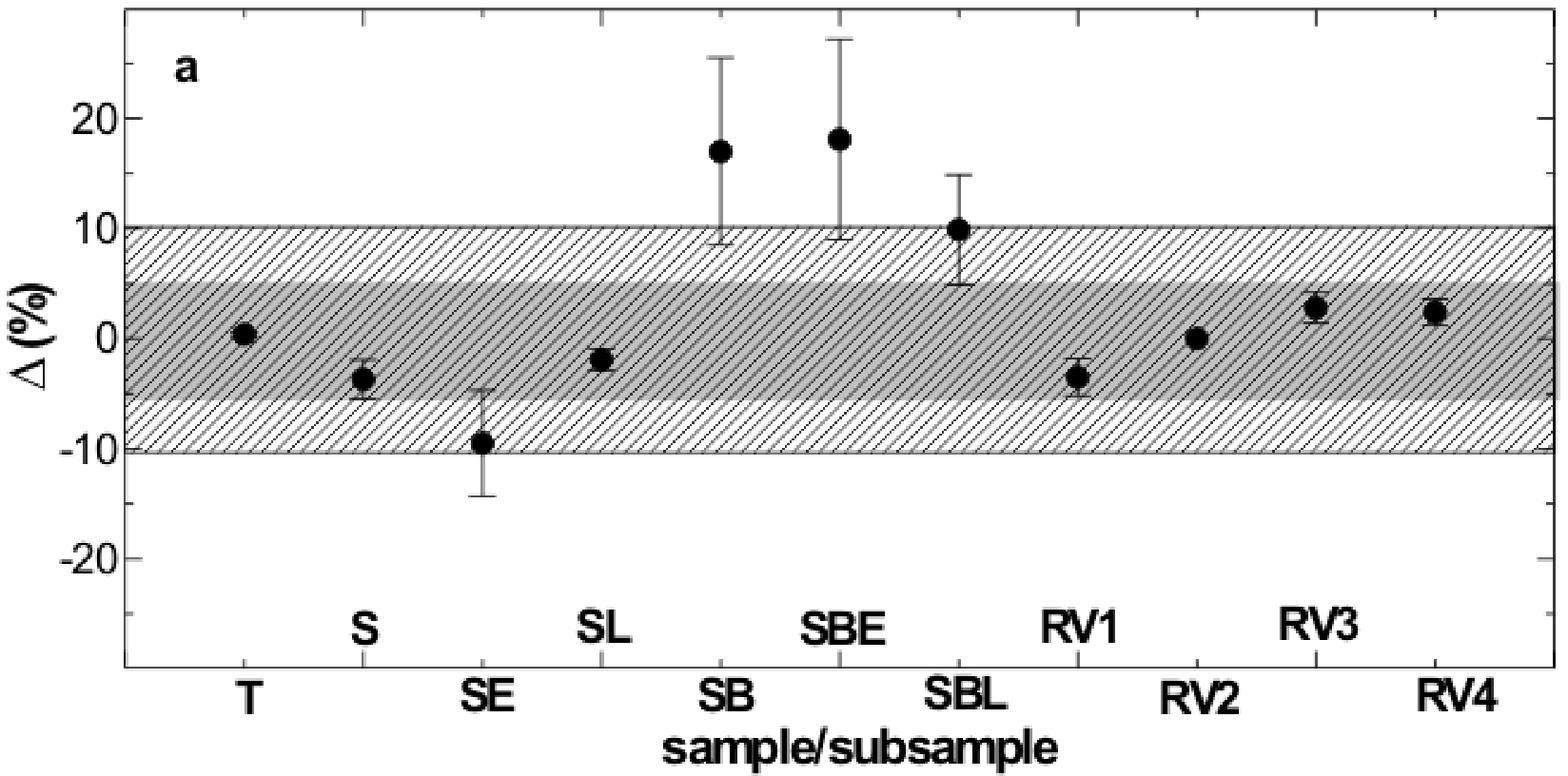}
      \includegraphics[height=3.55cm]{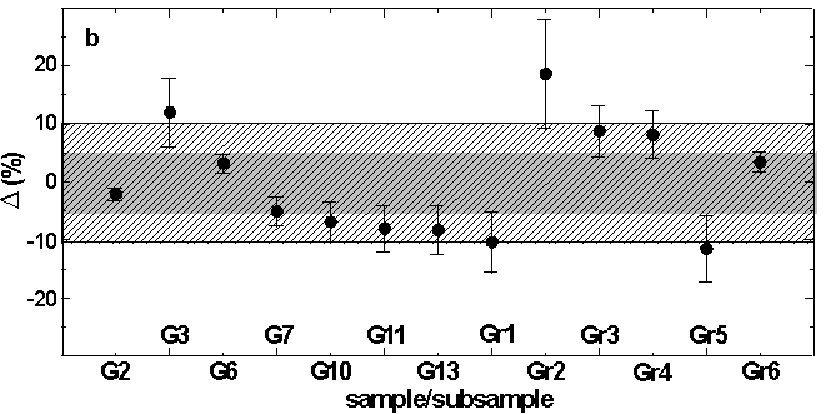}
      \caption[]{The basic statistics of the Z and S-mode of galaxies in the total sample and subsamples.
      The full form of the abbreviations (X-axis)
      are given in Table 2 (first column). $\Delta$(\%)=$S$$-$$Z$, where $S$
      and $Z$ represent the number of S- and Z-mode galaxies, respectively.
      The statistical error bars $\sigma$(\%) shown in the figure are calculated as: $\sigma$(\%) =
$\sigma$/($\sqrt{S}$+$\sqrt{Z}$)$\times$100, where $\sigma$ =
($\sqrt{S}$-$\sqrt{Z}$). The grey-shaded and the slanting-line
region represent the $\leq$$\pm$5\% and $\leq$$\pm$10\% $\Delta$
value, respectively. }
\end{figure}

\begin{table*}
      \caption[]{Statistics of leading (column 3) and trailing arm (column 4)
      galaxies in the total sample and subsamples. The fifth and sixth column give the numeral and percentage
      difference ($\Delta$ = $S$--$Z$) between the S- ($S$) and the Z- ($Z$) modes.
      The next two columns give the error: $\sigma$ = ($\sqrt{S}$--$\sqrt{Z}$) and
      $\sigma$(\%) = $\sigma$/($\sqrt{S}$+$\sqrt{Z}$)$\times$100. The eighth column gives the difference between the
      standard deviation (in arcmin) of the major diameters ($a$) of the S- and Z-modes galaxies ($\Delta$($a\,sde)$).
      The difference between the sum of the major diameters ($\Delta$($a$)\%) are listed in the last column.
      The sample/subsample and their abbreviations are given in first two columns.}
    $$
\begin{array}{p{0.30\linewidth}rccrrrrrr}
\hline\hline \noalign{\smallskip}
sample/subsample & $symbol$  & $Z$ & $S$ & $$\Delta$$ & $$\Delta$(\%)$ &  $$\sigma$(\%)$ & $$\Delta$($a$\,sde)$ & $$\Delta$($a$)(\%)$\\
\hline\hline \noalign{\smallskip}
Total                                 &   $T$     &   814 &   807 &   -7  &   -0.4    &  -0.2  &  0.019 &  0.0 \\
Spiral                                &   $S$     &   395 &   367 &   -28 &   -3.7    &   -1.8 &  0.031 &  3.0 \\
Spiral (early-type)                   &   $SE$    &   150 &   124 &   -26 &   -9.5    &   -4.8 &  0.058 &  8.1 \\
Spiral (late-type)                    &   $SL$    &   131 &   126 &   -5  &   -1.9    &   -1.0 &  0.031 &  0.3 \\
Barred Spiral                         &   $SB$    &   191 &   269 &   78  &   17.0    &   8.5  &  0.062 &  15.2 \\
Barred Spiral (early-type)            &   $SBE$   &   97  &   140 &   43  &   18.1    &   9.1  &  0.091 &  14.6 \\
Barred Spiral (late-type)             &   $SBL$   &   68  &   83  &   15  &   9.9     &   5.0  &  0.066 &  9.2 \\
3\,000$<$RV (km s$^{-1}$)$\leq$3\,500 &   $RV1$   &  208  &   194 &   -14 &   -3.5    &   -1.7 &  0.046 &  2.6 \\
3\,500$<$RV (km s$^{-1}$)$\leq$4\,000 &   $RV2$   &  201  &   201 &   0   &   0.0     &   0.0  &  0.031 &  3.1 \\
4\,000$<$RV (km s$^{-1}$)$\leq$4\,500 &   $RV3$   &  172  &   182 &   10  &   2.8     &   1.4  &  0.034 &  0.6 \\
4\,500$<$RV (km s$^{-1}$)$\leq$5\,000 &   $RV4$   &  226  &   237 &   11  &   2.4     &   1.2  &  0.041 &  1.0 \\
Grid 1                                &   $G1$    &  20   & 21    &   1   &   2.4     &   1.2   & 0.068 &  6.8 \\
Grid 2                                &   $G2$    &  121  & 116   &   -5  &   -2.1    &   -1.1  & 0.014 &  1.6 \\
Grid 3                                &   $G3$    &  88   & 112   &   24  &   12.0    &   6.0   & 0.076 &  9.3 \\
Grid 4                                &   $G4$    &  12   & 14    &   2   &   7.7     &   3.9   & 0.647 &  9.7 \\
Grid 5                                &   $G5$    &  11   & 8     &   -3  &   -15.8   &   -7.9  & 0.042 &  22.3 \\
Grid 6                                &   $G6$    &  75   & 80    &   5   &   3.2     &   1.6   & 0.081 &  4.5 \\
Grid 7                                &   $G7$    &  62   & 56    &   -6  &   -5.1    &   -2.5  & 0.095 &  8.9 \\
Grid 8                                &   $G8$    &  22   & 33    &   11  &   20.0    &   10.1  & 0.073 &  12.9 \\
Grid 9                                &   $G9$    &  20   & 31    &   11  &   21.6    &   10.9  & 0.028 &  18.1 \\
Grid 10                               &   $G10$   &  124  & 108   &   -16 &   -6.9    &   -3.5  & 0.004 &  5.4 \\
Grid 11                               &   $G11$   &  61   & 52    &   -9  &   -8.0    &   -4.0  & 0.025 &  6.2 \\
Grid 12                               &   $G12$   &  20   & 20    &   0   &   0.0     &   0.0   & 0.409 &  7.4 \\
Grid 13                               &   $G13$   &  78   & 66    &   -12 &   -8.3    &   -4.2  & 0.039 &  2.9 \\
Grid 14                               &   $G14$   &  37   & 44    &   7   &   8.6     &   4.3   & 0.356 &  10.3 \\
Grid 15                               &   $G15$   &  47   & 44    &   -3  &   -3.3    &   -1.6  & 0.050 &  5.2 \\
Grid 16                               &   $G16$   &  9    & 9     &   0   &   0.0     &   0.0   & 0.191 &  8.6 \\
Group 1                               &   $Gr1$   &  37   & 30    &   -7  &   -10.4   &   -5.2  & 0.032 &  7.1 \\
Group 2                               &   $Gr2$   &  48   & 70    &   22  &   18.6    &   9.4   & 0.097 &  12.6 \\
Group 3                               &   $Gr3$   &  31   & 37    &   6   &   8.8     &   4.4   & 0.027 &  4.3 \\
Group 4                               &   $Gr4$   &  34   & 40    &   6   &   8.1     &   4.1   & 0.031 &  3.9 \\
Group 5                               &   $Gr5$   &  107  & 85    &   -22 &   -11.5   &   -5.7  & 0.089 &  11.2 \\
Group 6                               &   $Gr6$   &  42   & 45    &   3   &   3.4     &   1.7   & 0.024 &  1.6 \\
\noalign{\smallskip} \hline\hline
\end{array}
     $$
\end{table*}

A significant dominance of S-mode galaxies are noticed
(17\%$\pm$8.5\%) in spiral barred galaxies. The $\Delta$ value is
found $>$ 9\% for both early- (SBa, SBab, SBb and SBbc) and
late-type (SBc, SBcd, SBd, SBm) barred spirals. Similar result
(i.e., $\Delta$ $>$ 8\%) is found for the irregulars and the
morphologically unidentified galaxies.

A similarity is noticed between the late-type spirals and barred
spirals: the $\Delta$ value for both the late-types are found to
be less than that of early-types (see Table 2).

The difference between the standard deviation of the major
diameters ($\Delta(a\,$sde$)$) for S- and Z-mode galaxies is found
less than 0.050 arc minute for the total sample, spirals and the
late-type spirals (eighth column, Table 1). These samples showed
$\Delta$ value $<$ 5\% (grey-shaded region, Fig. 2a). Thus, we
noticed a good correlation between the $\Delta$(\%) and
$\Delta(a\,$sde$)$ value.

The difference between the sum of the major diameters (in
percentage) are found greater than 10\% for the barred spirals and
early-type barred spirals. Interestingly, these two subsamples
showed $\Delta$ value greater than 15\% (Fig. 2a). Thus,
inhomogeniety in the distribution of S- and Z-mode galaxies is
noticed for barred spirals.

\subsubsection{Radial velocity}

A very good correlation between the number of S- and Z-mode can be
seen in the RV classifications (Fig. 1d). All 4 subsamples show
the $\Delta$ and $\Delta(a\,$sde$)$ values less than 5\% and
0.050, respectively (Table 2). In addition, $\Delta$($a$) is found
to be $<$ 5\%. This result is important in the sense that the
statistics in these subsamples is rich (number of galaxies $>$
170) enough. Thus, we could not observe preference structural
modes (S or Z) in the low and high RV galaxies in our database.

A difference is noticed: dominance of Z- and S- modes,
respectively in low (RV1) and high (RV3, RV4) RV subsamples.
However, this dominance is not significant (i.e., $\Delta$ $<$
5\%). An equal number of S- and Z-mode galaxies are found in the
subsample RV2 (3\,500 $<$ RV (km s$^{-1}$) $\leq$ 4\,000) (Table
1). In order to check the binning effect, we further classify the
total galaxies in 6 ($\Delta$RV = 333 km s$^{-1}$) and 8 bins
($\Delta$RV = 250 km s$^{-1}$) and study the statistics. No
significant dominance of either S- and Z-modes are noticed.

Thus, it is found that the homogeneous distribution of S- and
Z-mode galaxies remain invariant with the global expansion (i.e,
expansion of the Universe). We further discuss this result below.

\subsubsection{Area}

We study the distribution of S- and Z-mode galaxies by dividing
the sky into 16 equal parts (Fig. 3a). The area of the grid (G) is
90$^{\circ}$ $\times$ 45$^{\circ}$ (RA $\times$ Dec). The area
distribution of S- and Z-mode galaxies are plotted, that can be
seen in Fig. 3a'. The statistical parameters are given in Table 2.

\begin{figure}
\vspace{0.0cm}
      \centering
      \includegraphics[height=3.8cm]{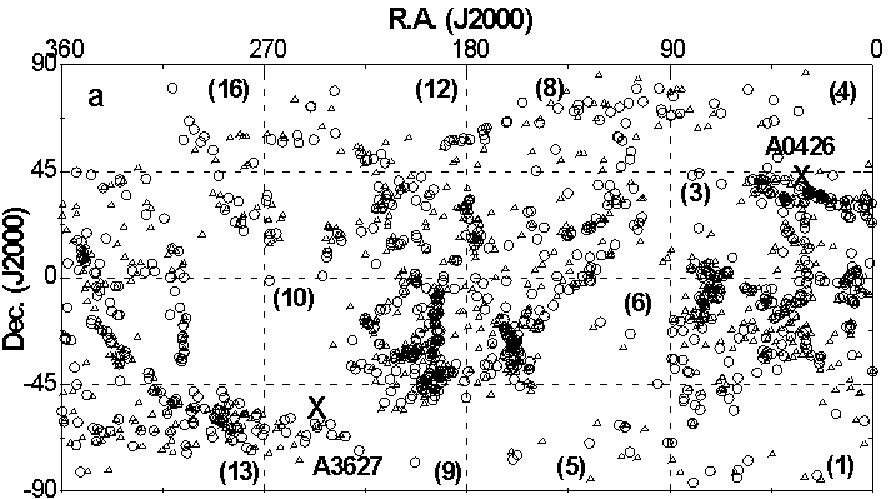}
      \includegraphics[height=3.8cm]{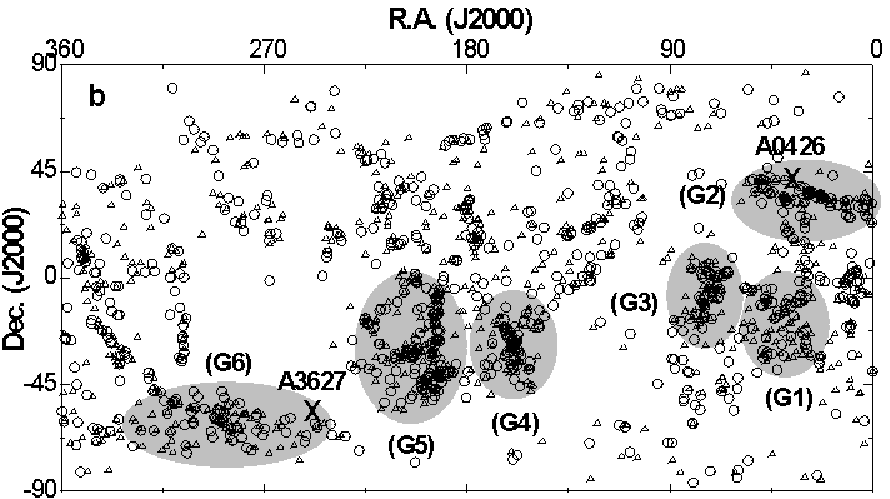}
      \includegraphics[height=2.5cm]{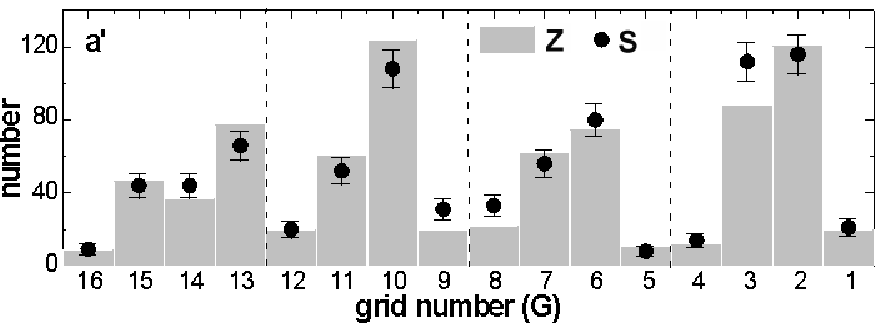}
      \includegraphics[height=2.5cm]{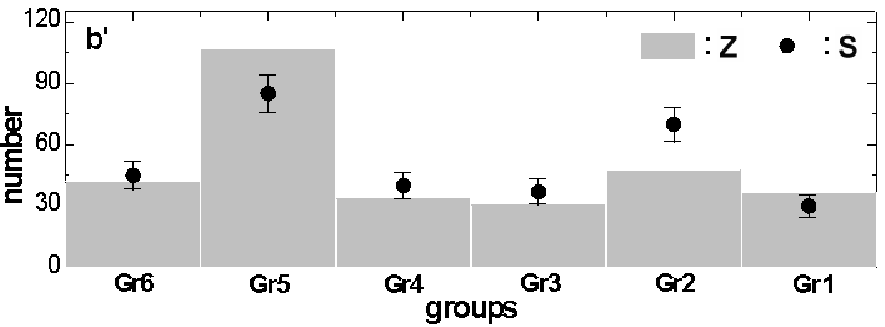}
      \caption[]{(a) All sky distribution of Z-mode (hollow circle) and S-mode (hollow triangle)
      galaxies in 16 area grids. (a') The histogram showing the
      distribution of the Z- and S-mode galaxies in the grids G1 to G16.
      (b) Six groups of the galaxies, represented by the grey-shaded region.
      (b') The distribution of Z- and S-modes in
      6 groups. The statistical error bar $\pm$1$\sigma$ is shown. The
positions of the clusters Abell 0426 and Abell 3627 are shown by
the symbol ``$\times$" (a,b).}
\end{figure}

A significant dominance ($>$2$\sigma$) of S-mode is noticed in
grid 3 (RA: 0$^{\circ}$ to 90$^{\circ}$, Dec: 0$^{\circ}$ to
45$^{\circ}$ (J2000)) (Fig. 3a,a'). An elongated group of galaxies
can be seen in this grid. In this grid, $\Delta$,
$\Delta(a\,$sde$)$ and $\Delta(a)\%$ are found to be 12\% $\pm$
6\%, 0.076 and 9.3\%, respectively. These figures suggest that the
distribution of S- and Z-mode galaxies in G3 is not homogeneous.
Probably, this is due to the apparent subgroupings or
subclusterings of the galaxies.

The S-mode galaxies dominate in the grids G8 and G9 (Fig. 3a').
However, the statistics is poor ($<$ 40) in these grids (Table 2).
In addition, no groupings or subclustering are noticed.

A dominance ($\sim$1.5$\sigma$) of Z-mode is noticed in G10 (RA:
180$^{\circ}$ to 270$^{\circ}$, Dec: --45$^{\circ}$ to 0$^{\circ}$
(J2000)) and G13 (RA: 270$^{\circ}$ to 360$^{\circ}$, Dec:
--90$^{\circ}$ to --45$^{\circ}$ (J2000)) (Fig. 3a,a'). In both
the grids, a large aggregation of the galaxies can be seen. A
subcluster-like aggregation can be seen in G10. An elongated
structure can be seen in G13. In both the grids, $\Delta$ value is
found to be greater than 5\% (Table 2).

No dominance of either S- and Z-mode galaxies is noticed in the
groups G1, G2, G4, G5, G6, G7, G11, G12, G14, G15 and G16. Thus,
homogeneous distribution of S- and Z-mode galaxies is found intact
in $\sim$ 80\% area of the sky. We suspect that the groupings or
subclusterings of the galaxies lead the preference structural
modes (S or Z) in G3, G10 and G13.

\subsubsection{Galaxy groups}

In all-sky map, several groups of galaxies can be seen (Fig. 3a).
It is interesting to study the distribution of structural modes (S
or Z) of galaxies in these groups. For this, we systematically
searched for the groups fulfilling following selection criteria:
(a) major diameter $>$ 30$^{\circ}$, (b) cutoff diameter $<$ 2
times the background galaxies, (c) number of galaxies $>$ 50. We
found 6 groups fulfilling these criteria (Fig. 3b). All 6 groups
(Gr) are inspected carefully. In 3 groups (Gr2, Gr5 and Gr6),
subgroups can be seen. The number of galaxies in the groups Gr2
and Gr5 are found more than 100.

The clusters Abell 0426 and Abell 3627 are located close to the
groups Gr2 and Gr6. The symbol ``$\times$" represents the position
of the cluster center in Fig. 3b. The mean radial velocities of
these clusters are 5\,366 km s$^{-1}$ and 4\,881 km s$^{-1}$,
respectively. However, we have removed the member galaxies of
these clusters from our database.

A significant dominance ($>$2$\sigma$) of S-mode galaxies is
noticed in the group Gr2 (Fig. 3b,b'). The $\Delta$,
$\Delta(a\,$sde$)$ and $\Delta(a)\%$ values are found to be 18.6\%
($\pm$9.4\%), 0.097 and 12.6\%, respectively (Table 2). We suspect
that the galaxies in this group is under the influence of the
cluster Abell 0426, due to which apparent subclustering of the
galaxies is seen.

The galaxies in Gr5 shows an opposite preference: a significant
dominance of the Z-mode galaxies ($>$2$\sigma$) (Fig. 3b,b'). In
this group, $\Delta$, $\Delta(a\,sde)$ and $\Delta(a)\%$ are found
to be 11.5\% $\pm$ 5.7\%, 0.089 and 11.2\%, suggesting
inhomogeneous distribution of structural modes (Table 2).

No humps or dips can be seen in the groups Gr1, Gr3, Gr4 and Gr6
(Fig. 3.2b,b'). Thus, the distribution of S- and Z-mode galaxies
in these groups are found to be homogeneous. The number of
galaxies in these groups are less than 100.

In the group 6, we could not notice the influence of the cluster
Abell 3627. This might be due to the off location of the cluster
center from the group center.

\subsection{Anisotropy in the position angle distribution}

\begin{figure} \vspace{0.0cm}
      \centering
      \includegraphics[height=3.4cm]{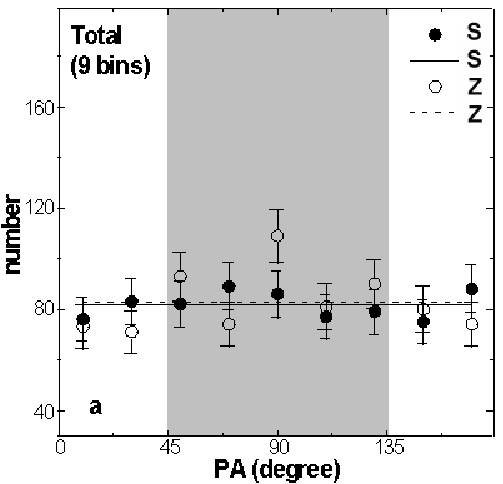}
      \includegraphics[height=3.4cm]{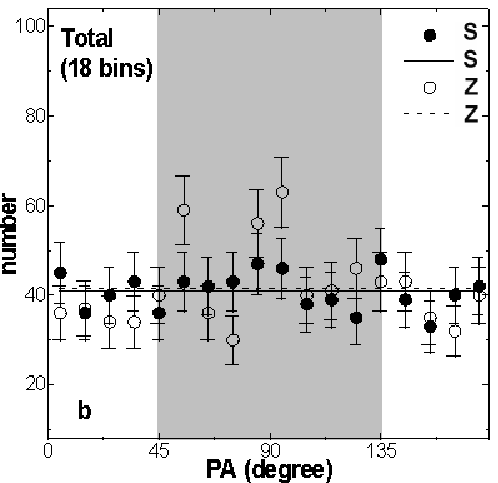}
      \caption[]{The equatorial position angle (PA) distribution of
      total Z- and S-mode galaxies plotted in 9 (a) and 18 (b)
      bins. The solid and the dashed line represent the expected
      isotropic distribution for S- and Z-mode
      galaxies, respectively.
      The observed counts with statistical $\pm$1$\sigma$ error bars are shown. PA =
      90$^{\circ}$$\pm$45$^{\circ}$ (grey-shaded region)
      corresponds to the galactic rotation axes tend to be oriented perpendicular
      with respect to the equatorial plane.}
\end{figure}

We study the equatorial position angle (PA) distribution of S- and
Z-mode galaxies in the total sample and subsamples. A spatially
isotropic distribution is assumed in order to examine non-random
effects in the PA-distribution. In order to discriminate the
deviation from the randomness, we use three statistical tests:
chi-square, auto correlation and the Fourier. The bin size was
chosen to be 20$^{\circ}$ (9 bins) in all these tests. The
statistically poor bins (number of solution $<$ 5) are omitted in
the analysis. The conditions for anisotropy are the following: the
chi-square probability P($>\chi^2$) $<$ 0.050, correlation
coefficient $C$/$\sigma(C)$ $>$ 1, first order Fourier coefficient
$\Delta_{11}$/$\sigma(\Delta_{11}$) $>$ 1 and the first order
Fourier probability P($>\Delta_1$)$<$0.150 as used by Godlowski
(1993). Table 3 lists the statistical parameters for the total
samples and subsamples.

In the Fourier test, $\Delta_{11}$ $<$ 0 (i.e., negative)
indicates an excess of galaxies with the galactic planes parallel
to the equatorial plane. In other words, a negative $\Delta_{11}$
suggests that the rotation axis of galaxies tend to be oriented
perpendicular with respect to the equatorial plane. Similarly,
$\Delta_{11}$ $>$ 0 (i.e., positive) indicates that the rotation
axis of galaxies tend to lie in the equatorial plane.

In the histograms (see Figs. 4-7), a hump at
90$^{\circ}$$\pm$45$^{\circ}$ (grey-shaded region) suggests that
the galactic planes of galaxies tend to lie in the equatorial
plane. In other words, the rotation axes of galaxies tend to be
oriented perpendicular with respect to the equatorial plane when
there is excess number of solutions in the grey-shaded region in
the histogram.

All three statistical tests show isotropy in the total S-mode
galaxies. Thus, no preferred alignment is noticed for the total
S-mode galaxies (solid circles in Fig. 4a). Interestingly, all
three statistical tests show anisotropy in the total Z-mode
galaxies. The chi-square and Fourier probabilities (P$(>\chi^2)$,
P($>\Delta_1$)) are found 1.5\% ($<$ 5\% limit) and 8.5\% ($<$
15\% limit), respectively (Table 2). The auto correlation
coefficient (C/C($\sigma$)) turned --3.2 ($>>$1). The
$\Delta_{11}$/$\sigma(\Delta_{11}$) value is found to be negative
at $\sim$ 2$\sigma$ level, suggesting that the rotation axes of
Z-mode galaxies tend to be oriented perpendicular the equatorial
plane. Three humps at 50$^{\circ}$ ($>$1.5$\sigma$), 90$^{\circ}$
($>$2$\sigma$) and 130$^{\circ}$ (1.5$\sigma$) support this result
(Fig. 4a). We checked the biasness in the results due to bin size
by increasing and decreasing the number of bins. A similar
statistical result is found for both structural modes. Fig. 4b
shows the PA-distribution histogram for the total sample in 18
bins. The Z-mode galaxies show three significant humps in the
grey-shaded region, supporting the results mentioned above.

Thus, we conclude isotropy for S-mode whereas anisotropy for
Z-mode galaxies in the total sample.
\begin{table*}
      \caption[]{Statistics of the PA-distribution of galaxies in the
      total sample and subsamples (first column).
The second, third, fourth and fifth columns give the chi-square
probability (P$(>\chi^2)$), correlation coefficient
(C/C($\sigma$)), first order Fourier coefficient
($\Delta_{11}$/$\sigma$($\Delta_{11}$)), and first order Fourier
probability P($>\Delta_1$), respectively. The last four columns
repeats the previous columns.}
    $$
         \begin{array}{p{0.1\linewidth}ccccccccc}
            \hline
            \noalign{\smallskip}
            sample  &   & $S-mode$ &    &  &   & $Z-mode$ &  &   \\
            \noalign{\smallskip}
                    &  $P$(>\chi^2)$$ & $C/C($\sigma$)$ & $$\Delta_{11}$/${\sigma}$($\Delta_{11}$)$ & $P(${>}\Delta_1$)$ & $P$(>\chi^2)$$ & $C/C($\sigma$)$ & $$\Delta_{11}$/${\sigma}$($\Delta_{11}$)$ & $P(${>}\Delta_1$)$ \\
            \hline
            \noalign{\smallskip}
            total   & 0.666 &  +0.0   & -0.9  & 0.381 & 0.015  & -3.2   & -1.9   & 0.085 \\
            S       & 0.511 &  -0.7   & -1.2  & 0.434 & 0.225  & +0.4   & +0.8   & 0.383 \\
            SE      & 0.973 &  +0.1   & -0.9  & 0.569 & 0.031  & +2.0   & +2.8   & 0.015 \\
            SL      & 0.234 &  +0.5   & +0.8  & 0.209 & 0.460  & -0.1   & -0.5   & 0.345 \\
            SB      & 0.729 &  +0.3   & +1.0  & 0.454 & 0.285  & -1.0   & -0.2   & 0.497 \\
            SBE     & 0.739 &  +0.1   & -0.5  & 0.566 & 0.230  & -0.7   & +0.1   & 0.521 \\
            SBL     & 0.043 &  +1.8   & +1.7  & 0.046 & 0.620  & -0.9   & -0.2   & 0.872 \\
            RV1     & 0.910 &  +0.3   & +0.8  & 0.362 & 0.369  & -0.9   & -0.6   & 0.285 \\
            RV2     & 0.790 &  +0.3   & -1.0  & 0.496 & 0.925  & -0.4   & -0.2   & 0.887 \\
            RV3     & 0.050 &  +1.6   & -1.5  & 0.083 & 0.033  & -1.8   & -2.3   & 0.046 \\
            RV4     & 0.043 &  -2.3   & -1.5  & 0.116 & 0.636  & +0.2   & -0.7   & 0.692 \\
            Gr2     & 0.455 &  +0.6   & +0.8  & 0.861 & 0.033  & -1.8   & +1.7   & 0.116 \\
            Gr5     & 0.033 &  -1.4   & -2.0  & 0.085 & 0.516  & +0.4   & -0.4   & 0.548 \\
            \noalign{\smallskip}
            \hline
         \end{array}
     $$
\end{table*}

\subsubsection{Morphology}


In the spirals, the chi-square and auto correlation tests show
isotropy for both the S- and Z-modes. The first order Fourier
probability is found greater than 35\%, suggesting no preferred
alignment. However, the $\Delta_{11}$ value exceeds 1$\sigma$
limit (--1.2$\sigma$) in the S-mode spirals. A hump at
90$^{\circ}$ is not enough to turn the
$\Delta_{11}$/$\sigma(\Delta_{11}$) $>$ 1.5 (Fig. 5a). Similarly,
a hump at 150$^{\circ}$ is not enough to make the
$\Delta_{11}$/$\sigma(\Delta_{11}$) $>$ 1.5 in the Z-mode spirals.
Hence, the preferred alignment is not profounded in both the S-
and Z-mode spirals. Thus, we conclude a random orientation of S-
and Z-mode spirals.

\begin{figure} \vspace{0.0cm}
      \centering
      \includegraphics[height=3.4cm]{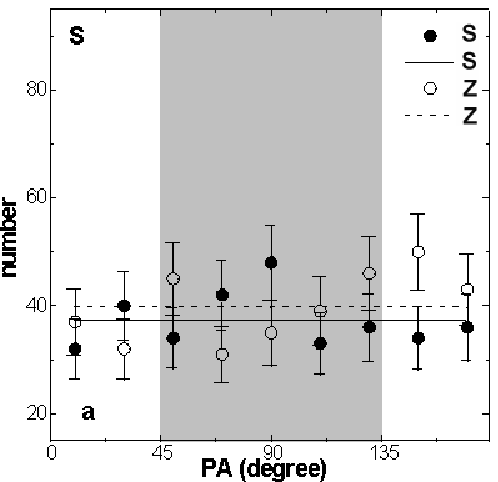}
      \includegraphics[height=3.4cm]{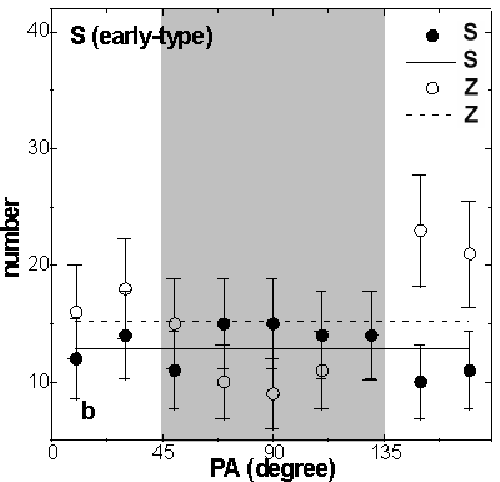}
      \includegraphics[height=3.4cm]{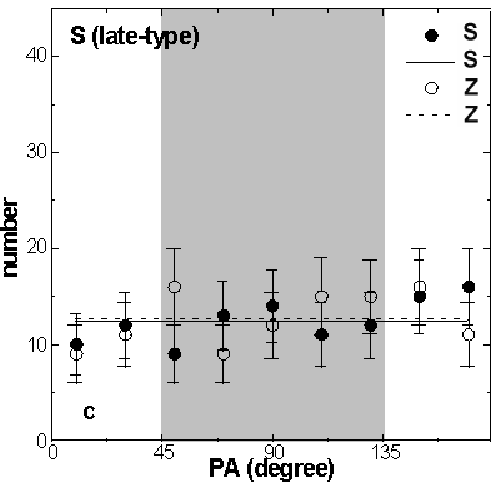}
      \includegraphics[height=3.4cm]{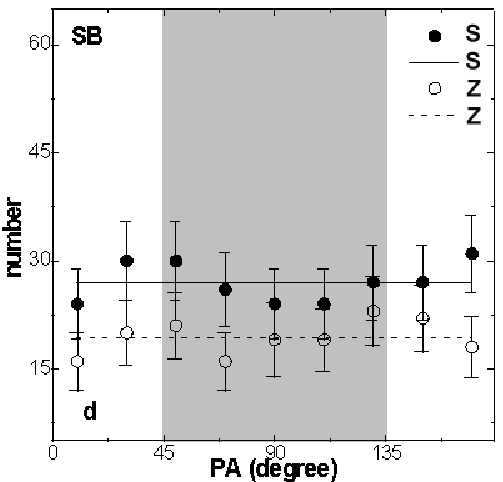}
      \includegraphics[height=3.4cm]{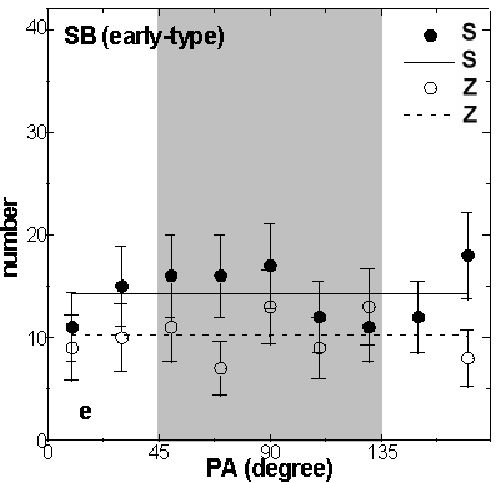}
      \includegraphics[height=3.4cm]{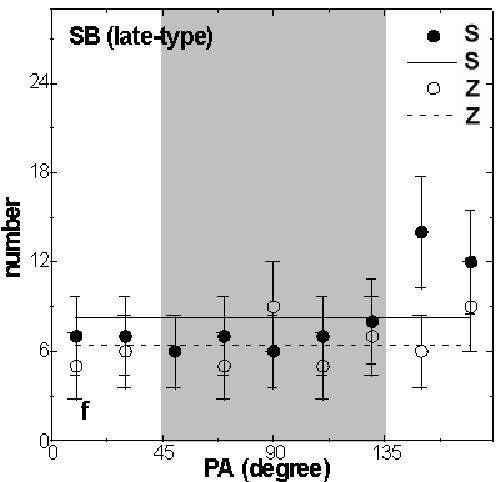}
      \caption[]{The equatorial PA-distribution of Z- and S-mode galaxies in the spirals (a),
      early-type spirals (b), late-type spirals (c), barred spirals (d),
      early-type barred spirals (e) and late-type barred spirals (f). The symbols, error
      bars, dashed lines and the explanations are analogous to Fig. 4.}
\end{figure}

Early- and late-type S-mode spirals show isotropy in all three
statistical tests (Table 3). No humps and the dips are seen in the
histograms (solid circles in Fig. 5b,c). Thus, the S-mode spirals
show a random alignment in the PA-distribution. In the subsample
SE, all three statistical tests show anisotropy (Table 2). Two
significant humps at $>$ 150$^{\circ}$ cause the first order
Fourier coefficient ($\Delta_{11}$) $>$ +2.5$\sigma$ (hollow
circle in Fig. 5b). Thus, a preferred alignment is noticed in the
early-type Z-mode spirals: the galactic rotation axes tend to lie
in the equatorial plane. The late-type Z-mode spirals show a
random alignment.

The spiral barred galaxies show a random alignment in both the S-
and Z-modes. In Fig. 5d, no deviation from the expected
distribution can be seen. All three statistical tests support this
result (Table 2). A similar result is found for the early-type SB
galaxies in both structural modes (Table 3, Fig. 5e).

The P$(>\chi^2)$ and P($>\Delta_1$) are found less than 5\%,
suggesting a preferred alignment for the late-type S-mode SB
galaxies (Table 3). The auto correlation coefficient
(C/C($\sigma$)) and the hump at $>$ 150$^{\circ}$ support this
result (Fig. 5f). The $\Delta_{11}$/$\sigma(\Delta_{11}$) is found
to be positive at 1.7$\sigma$ level, suggesting that the S-mode
SBL galaxies tend to lie in the equatorial plane. Thus, the
late-type S- and Z-mode SB galaxies show preferred and random
alignments, respectively.

\subsubsection{Radial velocity}

The subsamples RV1 and RV2 show isotropy in all three statistical
tests (Table 2). No humps or dips can be seen in Figs. 6a,b. Thus,
the galaxies having radial velocity in the range 3\,000 km
s$^{-1}$ to 4\,000 km s$^{-1}$ show a random alignment for both
the S- and Z-mode galaxies.

The humps at 90$^{\circ}$ ($>$2$\sigma$) and 110$^{\circ}$
($>$2$\sigma$) are found in the Z- and S-mode RV3 galaxies,
respectively (Fig. 6c). These two significant humps lead the
subsample to show anisotropy in the statistical tests (Table 2).
The $\Delta_{11}$ values are found negative at $\geq$1.5 level,
suggesting a similar preferred alignment for both modes: the
galaxy rotation axes tend to be directed perpendicular to the
equatorial plane.

\begin{figure} \vspace{0.0cm}
      \centering
      \includegraphics[height=3.4cm]{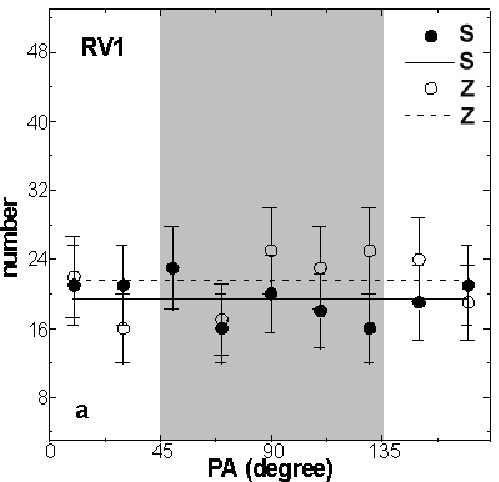}
      \includegraphics[height=3.4cm]{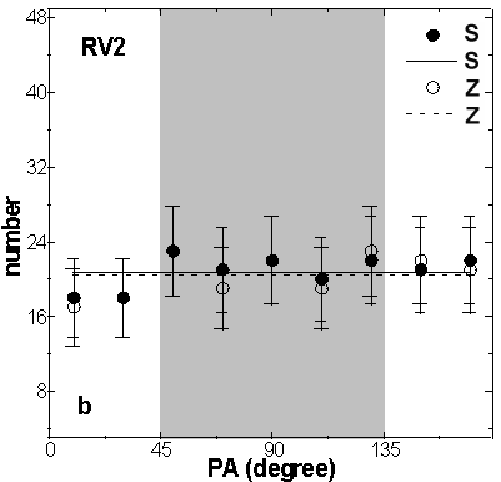}
      \includegraphics[height=3.4cm]{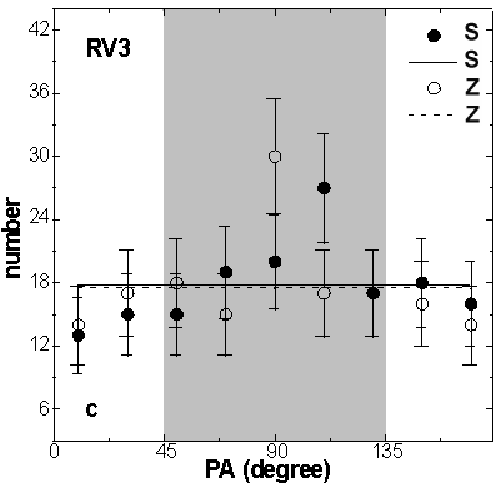}
      \includegraphics[height=3.4cm]{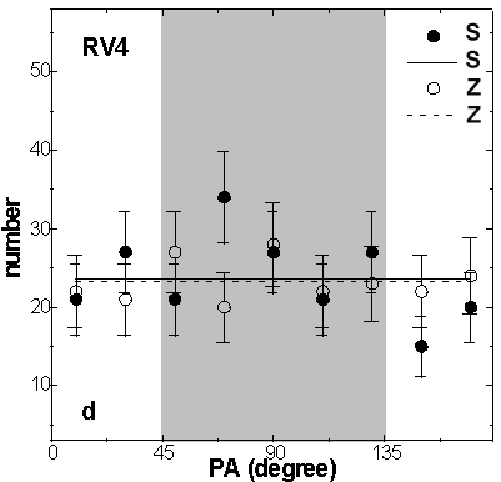}
      \caption[]{The equatorial PA-distribution of Z- and S-mode galaxies in RV1 (a), RV2 (b), RV3 (c) and RV4 (d). The abbreviations are listed
      in Table 1. The symbols, error bars, dashed lines and the explanations are analogous to Fig. 4.}
\end{figure}

A hump at 70$^{\circ}$ ($>$1.5$\sigma$) and a dip at 150$^{\circ}$
($\sim$2$\sigma$) cause the S-mode RV4 galaxies to show anisotropy
in all three statistical tests (Fig. 6d). Thus, the S-mode
galaxies having radial velocity in the range 4\,500 km s$^{-1}$ to
5\,000 km s$^{-1}$ show a similar preferred alignments as shown by
the subsample RV3: galactic planes of galaxies tend to lie in the
equatorial plane. The leading arm galaxies in the subsample RV4
show a random alignment (Table 3, Fig. 6d).

\subsubsection{Groups}

We do not study PA-distribution of S- and Z-mode galaxies in the
groups Gr1, Gr3, Gr4 and Gr6 because of poor statistics (number
$<$ 50).

We study the PA-distribution of S- and Z-mode galaxies in the
groups Gr2 and Gr5, where the dominance of either Z- or S-mode is
noticed. In addition, the statistics is relatively better in these
groups.
\begin{figure} \vspace{0.0cm}
      \centering
      \includegraphics[height=3.4cm]{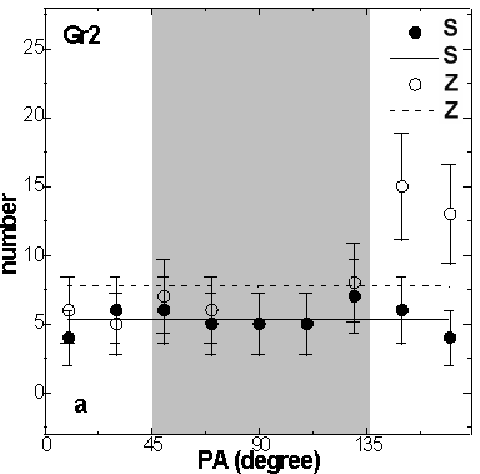}
      \includegraphics[height=3.4cm]{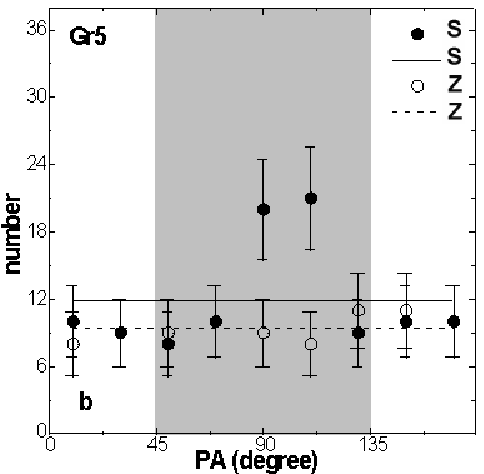}
      \caption[]{The equatorial PA-distribution of Z- and S-mode galaxies in the groups Gr2 and Gr5. The abbreviations are listed
      in Table 1. The symbols, error bars, dashed lines and the explanations are analogous to Fig. 4.}
\end{figure}

In the group Gr2, Z-mode galaxies dominate the S-mode galaxies. In
this group, the Z-mode galaxies show a preferred alignment whereas
S-mode galaxies show a random alignment in the PA-distribution.
All three statistical tests suggest anisotropy in the Z-mode
galaxies (Table 3). The humps at $>$ 150$^{\circ}$ cause the
$\Delta_{11}$ value to be positive at $>$ 1.5$\sigma$ level (Fig.
7a), suggesting that the rotation axes of Z-mode galaxies in Gr2
tend to be oriented parallel the equatorial plane.

The S-mode galaxies dominate in the group Gr5. Interestingly, a
preferred alignment of S-mode galaxies is noticed in the
PA-distribution. In Fig. 7b, two significant humps at 90$^{\circ}$
($\sim$2$\sigma$) and 110$^{\circ}$ ($>$2$\sigma$) can be seen.
These humps lead the subsample (S-mode Gr5) to show anisotropy in
the statistical tests (Table 3). No preferred alignment is noticed
in the Z-mode galaxies of this group.

Thus, the dominating structural modes (Z or S) show a preferred
alignment in the PA-distribution. This is noticeable result.

\subsection{Discussion}

Fig. 8 shows a comparison between the number ($\Delta$) and
position angle ($\Delta_{11}$/$\sigma(\Delta_{11}$)) distribution
of S- and Z-mode galaxies in the total sample and subsamples. This
plot deals the correlation between the homogeneity in the
structural modes and the random alignment in the subsamples. The
grey-shaded region represents the region of isotropy and
homogeneity for the $\Delta_{11}$/$\sigma(\Delta_{11}$) and
$\Delta$(\%), respectively.
\begin{figure}
\vspace{0.0cm}
      \centering
       \includegraphics[height=4cm]{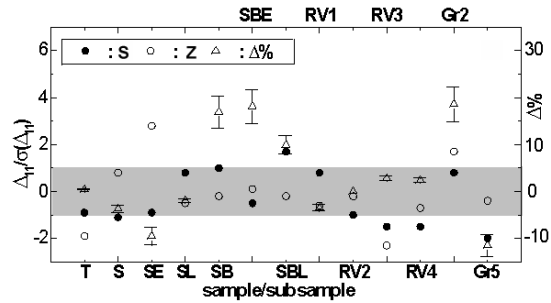}
      \caption[]{A comparison between the number ($\Delta$\%) and
the position angle ($\Delta_{11}$/$\sigma(\Delta_{11}$))
distribution of Z- and S-mode galaxies in the total sample and
subsamples.}
\end{figure}
Twenty five (out of 39, 64\%) subsamples lie in the grey-shaded
region (Fig. 8a), suggesting a good agreement between the
homogeneous distribution of S- and Z-mode galaxies and the random
alignment of the rotation axes of galaxies. In four subsamples
(SE, SBL, Gr2 and Gr5), a good correlation between the preferred
alignment and the dominance of either S- or Z-mode galaxies is
noticed (Fig. 8a). Thus, it is noticed that the random alignment
of the PAs of galaxies hint the existence of inhomogeneity in the
structural modes.

Aryal \& Saurer (2006) and Aryal, Paudel \& Saurer (2007) studied
the spatial orientation of galaxies in 32 Abell clusters of BM
type I, II, II-III and III and found a significant preferred
alignment in the late-type cluster (BM type II-III, III). They
concluded that the randomness decreases systematically in galaxy
alignments from early-type (BM type I, II) to late-type (BM type
II-III, III) clusters.

We noticed a very good correlation between the random alignments
and the homogeneity in the structural modes. Probably, this result
reveals the fact that the progressive loss of homogeneity in the
structural modes might have some connection with the rotationally
supported (spirals, barred spirals) to the randomized
(lenticulars, ellipticals) system. Thus, we suspect that the
dynamical processes in the cluster evolution (such as late-type
clusters) give rise to a dynamical loss of homogeneity in the
structural modes. It would be interesting to test this prediction
by studying the S- and Z-type spirals in the late-type clusters in
the future.

As 60\% of galaxies in the nearby universe are rotationally
supported discs, understanding angular-momentum acquisition is
obviously a crucial part of understanding galaxy evolution. The
winding sense of spiral arm patterns (morphological feature)
allows us to infer the orientation of the angular-momentum vector
of the disc galaxy. The expected distribution of spin vectors of
galaxies shows markedly different trends according to the galaxy
formation scenarios. One can suspect the possibility that the
actual distribution of galaxy spin shows a dipole or a quadrupole
component depending on the scenarios of galaxy formation. If
galaxy spins were generated according to the primordial whirl
scenario, a strong bias in either the S or Z patterns would be
seen in a face-on sample of galaxies. We did not find this trend
in our sample. A quadrupole distribution of S/Z might be observed
if the primary process was the generation of spins due to the
pancake shock scenario or the explosion scenario. On the other
hand, if the galaxy spins were produced by the tidal spin-up
process, there would be no global anisotropy as we noticed in many
cases, unless galaxy-cluster tidal interaction rather than
galaxy-galaxy tidal interaction were the primary process. No
significant correlation, however, was identified in any ensemble.

\section{Conclusions}
\label{sect:conclusion} We studied the winding sense of 1\,621
field galaxies around the Local Supercluster.  These galaxies have
radial velocity (RV) in the range 3\,000 km s$^{-1}$ to 5\,000 km
s$^{-1}$. The distribution of Z- and S-mode galaxies is studied in
the total sample and 32 subsamples. To examine non-random effects,
the equatorial position angle (PA) distribution of galaxies in the
total sample and subsamples are studied. In order to discriminate
anisotropy from the isotropy we have performed three statistical
tests: chi-square, auto-correlation and the Fourier. Our results
are as follows:

\begin{enumerate}

\item The homogeneous distribution of the total Z- and S-mode
galaxies is found, suggesting the homogeneous distribution of
winding sense (S or Z) of galaxies having RVs 3\,000 km s$^{-1}$
to 5\,000 km s$^{-1}$. The PA-distribution of S-mode galaxies is
found to be random, whereas preferred alignment is noticed for
Z-mode galaxies. It is found that the galactic rotation axes of
Z-mode galaxies tend to be oriented perpendicular the equatorial
plane.

\item Z-mode are found 3.7\% ($\pm$1.8\%) more than that of the
S-mode in the spirals whereas a significant dominance (17\% $\pm$
8.5\%) of S-mode is noticed in the barred spirals. This difference
is found $>$ 8\% for the irregulars and the morphologically
unidentified galaxies. A random alignment is noticed in the
PA-distribution of Z- and S-mode spirals. Thus, it is noticed that
the random alignment of the PAs of galaxies lead the existence of
inhomogeneity in the structural modes of galaxies.

\item The inhomogeneity in the structural modes is found stronger
for the late-type spirals (Sc, Scd, Sd and Sm) than that of
early-type (Sa, Sab, Sb and Sbc). Similar result is found for the
late-type barred spirals.

\item A very good correlation between the number of Z- and S-mode
galaxies are found in the RV subsamples. All 4 subsamples show the
$\Delta$ value less than 5\%. Thus, we conclude that the
homogeneous distribution of structural modes of field galaxies
remain invariant with the global expansion.

\item The galaxies having RVs 3\,000 km s$^{-1}$ to 4\,000 km
s$^{-1}$ show a random alignment for both the Z- and S-modes. The
rotation axes of Z- and S-mode galaxies having 4\,000 $<$ RV (km
s$^{-1}$) $\leq$ 4\,500 tend to be oriented perpendicular the
equatorial plane.

\item The distribution of the winding sense of galaxies is found
homogeneous in $\sim$ 80\% area of the sky. This property is found
to be violated in few groups of galaxies. Two such groups (Gr2 and
Gr8) are identified. In these groups, the structural dominance and
the preferred alignments of galaxies are found to oppose each
other.

\end{enumerate}

The true structural mode of a galaxy must involve a determination
of which side of the galaxy is closer to the observer (Binney and
Tremaine 1987). Three-dimensional determination of the leading and
the trailing arm patterns in the galaxies is a very important
problem. We intend to address this problem in the future.

\begin{acknowledgements}
We are indebted to the referee for his/her constructive criticism
and useful comments. I acknowledge Profs. R. Weinberger and W.
Saurer of Innsbruck University, Austria for insightful
discussions. I am thankful to Tribhuvan University, Nepal and
Innsbruck University, Austria for providing financial assistance
to visit Innsbruck University during Jan-March 2009.
\end{acknowledgements}



\label{lastpage}

\end{document}